\documentclass[12pt]{article}
\usepackage{epsf}
\title{ {\bf Neutral Higgs  $H^0 \rightarrow h^0 (A^0)\, l_i^- l_j^+$
decay }}

\author{\vspace{1cm}\\
        {\bf E. O. Iltan}
        \thanks{E-mail address:
        eiltan@heraklit.physics.metu.edu.tr}
 \\
        Physics Department, Middle East Technical University \\
        Ankara, Turkey\\}
\date{}
\begin{document}
\setlength{\baselineskip}{24pt}
\maketitle
\setlength{\baselineskip}{7mm}
\begin{abstract}
We study the lepton flavor violating $H^0 \rightarrow h^0 (A^0)\,
l_i^- l_j^+$ and the lepton flavor conserving $H^0 \rightarrow h^0
(A^0)\, l_i^- l_i^+$ $(l_i=\tau, l_j=\mu)$ decays in the framework
of the general two Higgs doublet model, the so-called model III.
We estimate the decay width of LFV (LFC) process at the order of
magnitude of $10^{-5}\, GeV$ ($10^{-4} \,GeV$), for $m_{H^0}=
150\, GeV$, and the intermediate values of the coupling
$\bar{\xi}^{E}_{N,\tau \mu}\sim \,5 \,GeV$ ($\bar{\xi}^{E}_{N,\tau
\tau}\sim \,30\, GeV$). The experimental result of the process
under consideration can ensure comprehensive information about the
standard model Higgs boson $H^0$ and the model III free
parameters.
\end{abstract}
\thispagestyle{empty}
\newpage
\setcounter{page}{1}
\section{Introduction}

The standard model (SM) of electroweak interactions is based on
the existence of the CP even Higgs boson ($H^0$). Its possible
detection in future collider experiments will ensure a test for
the SM and the strong information about the mechanism of the
electroweak symmetry breaking, the Higgs mass and the top Higgs
Yukawa coupling . The search for the Higgs boson is one of the
prime goals of LHC.

The light Higgs boson, $m_{H^0} \leq 130 \, GeV$, mainly decays
into $b \bar{b}$ pair \cite{Djouadi}. However its detection is
difficult due to the QCD background and the $t\bar{t} H^0$
channel, where the Higgs boson decays to  $b \bar{b}$, is the most
promising one \cite{Drollinger}. The supersymmetric QCD
contribution to  $H^0 b \bar{b}$ has been analyzed at one loop in
the Minimal Supersymmetric Model (MSSM) in the decoupling limit in
\cite{Howard}.

For a heavier Higgs boson $m_{H^0} \sim 180\, GeV$, the suitable
production exist via gluon fusion and the leading decay mode is
$H^0\rightarrow W W \rightarrow l^+ l'^- \nu_l \nu_{l'}$
\cite{RunII,Dittmar}. In\cite{Dittmar}, it is stated that this
decay mode gives three order times larger events compared to the
mode $H^0\rightarrow Z Z^* \rightarrow l^+ l^- l'^+ l'^-$. In this
work two new cuts have been employed to separate the irreducible
continuum background production. The decay $H^0\rightarrow A^0
A^0$ has been studied in the minimal supersymmetric model (MSSM)
including one-loop corrections and the decay width is obtained at
the order of magnitude of $10^{-2} GeV$.

The flavor violating (FV) interactions are important in the sense
that they do not exist in the SM and their analysis ensure
comprehensive information about the physics beyond the SM. The
lepton FV (LFV) $H^0$ decays have been studied in series of works
in the literature. $H^0\rightarrow \tau\mu$ decay is an example
for LFV decays and it has been studied in \cite{Cotti, Marfatia}.
In \cite{Cotti} a large $BR$, at the order of magnitude of
$0.1-0.01$, has been estimated in the framework of the 2HDM. In
\cite{Marfatia} its BR was obtained in the interval $0.001-0.01$
for the Higgs mass range $100-160 (GeV)$, for the LFV parameter
$\kappa_{\mu\tau}=1$. The work in \cite{Koerner} was due to the
observable CP violating asymmetries in the leptonic flavor
changing $H^0$ decays with branching ratios (BR s) of the order of
$10^{-6}-10^{-5}$. The LFV $H^0\rightarrow l_i l_j$ decay has been
studied also in \cite{Assamagan}.

In our work we analyze the LFV $H^0 \rightarrow h^0 l_i^- l_j^+$,
$H^0 \rightarrow A^0 l_i^- l_j^+$ and the lepton flavor conserving
(LFC) $H^0 \rightarrow h^0 l_i^- l_i^+$, $H^0 \rightarrow A^0
l_i^- l_i^+$ $(l_i=\tau, l_j=\mu)$ decays in the framework of the
general 2HDM, the so-called model III. These processes can exist
at the tree level in the model III and provide wide information
about the free parameters of the model, since their decay widths
depend on the masses of the new particles, namely $m_{h^0},
m_{A^0}$ and the leptonic Yukawa couplings. In our analysis, we
observe the decay width at the order of magnitude of $10^{-5}\,
GeV$, for outgoing $\tau$ and $\mu$ leptons, for the LFV process.
In the case of LFC decay, the decay width reaches of
$10^{-3\dot{}}\, GeV$, for the appropriate choice of the free
parameters, for outgoing $\tau \, \tau$ leptons. These numbers are
sensitive to the Yukawa coupling for $\tau-\mu$ ($\tau-\tau$)
transitions, the new Higgs boson masses $m_{h^0}$ and $m_{A^0}$,
and the SM Higgs mass $m_{H^0}$. Therefore their experimental
investigations will provide a crucial information about the SM
Higgs and the model III free parameters.

The paper is organized as follows: In Section II, we present the
theoretical expression for the decay widths of the LFV decay $H^0
\rightarrow h^0 l_i^- l_j^+$, $H^0 \rightarrow A^0 l_i^- l_j^+$
and the LFC decay $H^0 \rightarrow h^0 l_i^- l_i^+$, $H^0
\rightarrow A^0 l_i^- l_i^+$ $(l_i=\tau, l_j=\mu)$, in the
framework of the model III. Section 3 is devoted to discussion and
our conclusions.
\section{$H^0 \rightarrow h^0 (A^0)\, l_i^- l_j^+$, decays in the two
Higgs doublet model}
The decay of the SM Higgs $H^0$ into the new neutral Higgs bosons
$h^0$ ($A^0$) and the lepton pairs is kinematically allowed. In
the case of different flavors as an output in the lepton sector
the tree level LFV interactions should be to switch on. For such
interactions the model III version of the 2HDM is the simplest
candidate. In the model III, the Yukawa lagrangian, including the
interaction between the scalar bosons and the leptons, reads
\begin{eqnarray}
{\cal{L}}_{Y}= \eta^{E}_{ij} \bar{l}_{i L} \phi_{1} E_{j R}+
\xi^{E}_{ij} \bar{l}_{i L} \phi_{2} E_{j R} + h.c. \,\,\, ,
\label{lagrangian}
\end{eqnarray}
where $i,j$ are family indices of leptons, $L$ and $R$ denote
chiral projections $L(R)=1/2(1\mp \gamma_5)$, $\phi_{i}$ for
$i=1,2$, are the two scalar doublets, $l_{i L}$ and $E_{j R}$ are
lepton doublets and singlets respectively.
In general, there is a mixing between the neutral CP even bosons,
$H^0$ and $h^0$, however, by considering the gauge and $CP$ invariant
Higgs potential which spontaneously breaks  $SU(2)\times U(1)$ down to
$U(1)$  as
\begin{eqnarray}
V(\phi_1, \phi_2 )&=&c_1 (\phi_1^+ \phi_1-v^2/2)^2+ c_2 (\phi_2^+
\phi_2)^2 \nonumber \\ &+&  c_3 [(\phi_1^+ \phi_1-v^2/2)+ \phi_2^+
\phi_2]^2 + c_4 [(\phi_1^+ \phi_1)
(\phi_2^+ \phi_2)-(\phi_1^+ \phi_2)(\phi_2^+ \phi_1)] \nonumber \\
&+& c_5 [Re(\phi_1^+ \phi_2)]^2 + c_{6} [Im(\phi_1^+ \phi_2)]^2
+c_{7} \, , \label{potential}
\end{eqnarray}
with constants $c_i, \, i=1,...,7$, and the doublets $\phi_{1}$ and
$\phi_{2}$ as
\begin{eqnarray}
\phi_{1}=\frac{1}{\sqrt{2}}\left[\left(\begin{array}{c c}
0\\v+H^{0}\end{array}\right)\; + \left(\begin{array}{c c}
\sqrt{2} \chi^{+}\\ i \chi^{0}\end{array}\right) \right]\, ;
\phi_{2}=\frac{1}{\sqrt{2}}\left(\begin{array}{c c}
\sqrt{2} H^{+}\\ H_1+i H_2 \end{array}\right) \,\, ,
\label{choice}
\end{eqnarray}
where only $\phi_{1}$ has a vacuum expectation value;
\begin{eqnarray}
<\phi_{1}>=\frac{1}{\sqrt{2}}\left(\begin{array}{c c}
0\\v\end{array}\right) \,  \, ; <\phi_{2}>=0 \,\, ,
\label{choice2}
\end{eqnarray}
this mixing is switched off. Therefore,  $H_1$ and $H_2$ are obtained
as the mass eigenstates $h^0$ and $A^0$, respectively.

Now, we consider the lepton flavor violating processes $H^0
\rightarrow h^0 l_i^- l_j^+$ and $H^0 \rightarrow A^0 l_i^- l_j^+$
where $l_i,\, l_j$ are different leptons flavors, $e,\mu, \tau$
(see Fig. 1). The source of the LFV interactions are the Yukawa
couplings $\xi^{E}_{ij}$, which are responsible for the tree level
$h^0 (A^0)-l_i-l_j$ interactions. These couplings are complex in
general and they are the free parameters of the model III version
of 2HDM. Notice that, in the following, we replace $\xi^{E}$ with
$\xi^{E}_{N}$ where 'N' denotes the word 'neutral'.

\begin{figure}[htb]
\vskip 0.0truein \centering \epsfxsize=6.8in
\leavevmode\epsffile{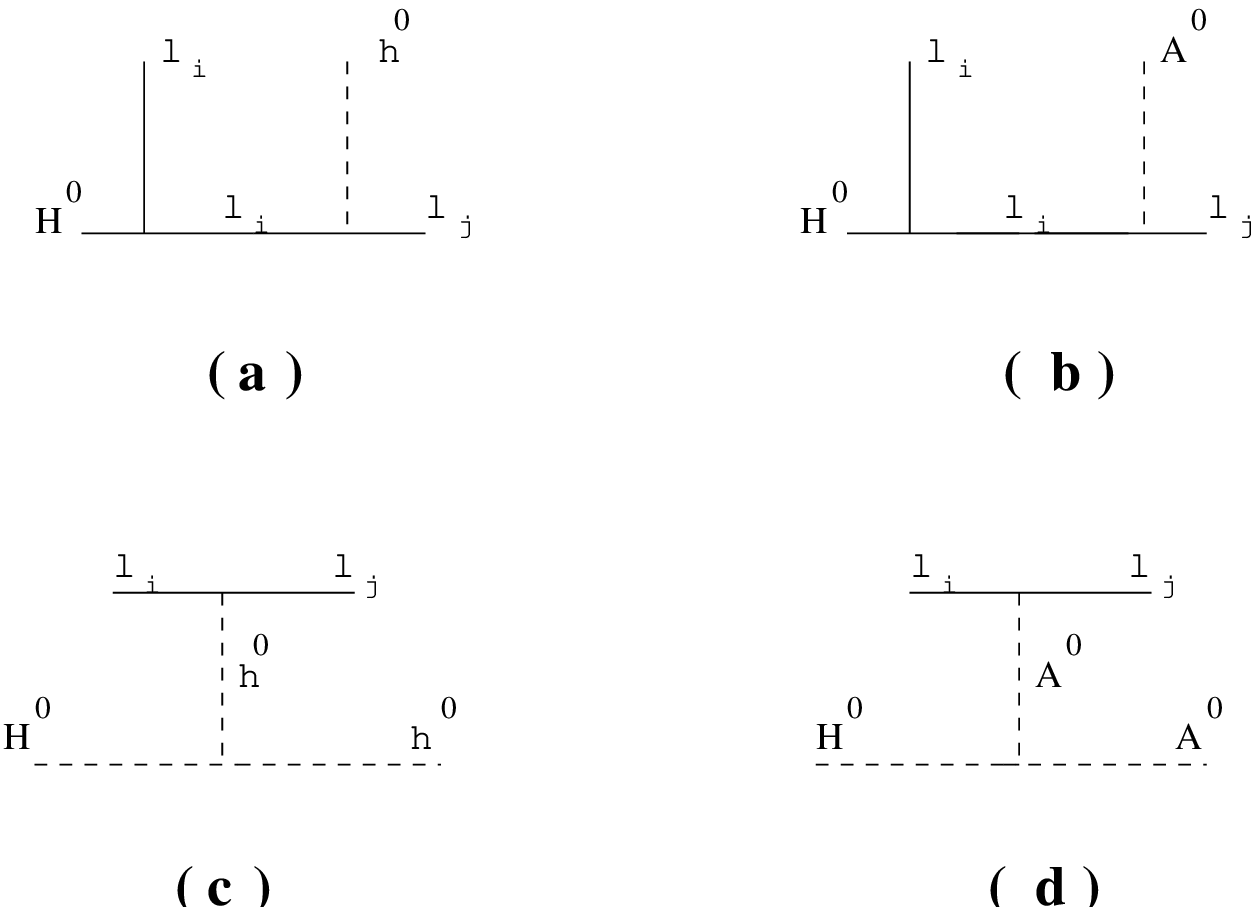} \vskip 1.0truein \caption[]{Tree
level diagrams contribute to $\Gamma (H^0\rightarrow h^0 (A^0)\,
l_i^- l_j^+)$, $i=e,\mu,\tau$ decay  in the model III version of
2HDM. Solid lines represent leptons, dashed lines represent the
$H^0$, $h^0$ and $A^0$ fields.} \label{fig1}
\end{figure}

Using the diagram given in Fig. \ref{fig1}, the matrix element
square of the process $H^0 \rightarrow h^0 l_i^- l_j^+$ is
obtained as
\begin{eqnarray}
|M|^2=\frac{g^4}{4 m^4_{W}}(A_1+A_2+A_3) \nonumber \\
\label{M2h0ij}
\end{eqnarray}
where
\begin{eqnarray}
 A_1&=&\frac{1}{2 (m^2_{H^0}+2 p.k_{l_i})^2}\, \, \Bigg\{
m^2_{l_i}\, |\bar{\xi}^{E}_{N,j i}|^2 \, \Bigg( 2
(p.k_{l_i})^2+(m^2_{H^0}- 4
m^2_{l_i})\,q.k_{l_i}+p.k_{l_i}\,(m^2_{H^0}\nonumber \\ &+& 4
m_{l_i}(m_{l_j}+2\,m_{l_i}-2 m_{l_j}\, sin^2\theta_{ij}) - 2 p.q)+
m_{l_i}(4\, m^2_{l_i}(m_{l_i}+m_{l_j}-2\,m_{l_j}\,
sin^2\theta_{ij})\nonumber
\\ &+& m^2_{H^0} (3\,m_{l_i}+m_{l_j}-2\,m_{l_j}\,
sin^2\theta_{ij})-4\,m_{l_j}\, p.q \Bigg)\Bigg\} \, , \nonumber
\\
A_2&=&\frac{1}{(m^2_{H^0}+2 p.k_{l_i})}\, \, \Bigg\{ 2\,
m_{l_i}\,m^2_{h^0}\, |\bar{\xi}^{E}_{N,j i}|^2 \,Im[p_{h^0}]
\Bigg( (3 m_{l_i}+m_{l_j}-2\,m_{l_j}\,sin^2\theta_{ij})\,p.k_{l_i}
\nonumber
\\ &+& m_{l_i}\,(m^2_{H^0}+ 2\, m^2_{l_i}+2 m_{l_i} m_{l_j}-4\,
m_{l_i} m_{l_j}\,sin^2\theta_{ij} -2 \,q.(k_{l_i}-p))\Bigg)
\Bigg\}\, , \nonumber
\\
A_3&=& 2 m^4_{h^0}\, |\bar{\xi}^{E}_{N,j i}|^2 \,Abs[p_{h^0}]^2
\Bigg(
m_{l_i}(m_{l_i}+m_{l_j}-2\,m_{l_j}\,sin^2\theta_{ij})+(p-q).k_{l_i}
\Bigg) \, . \label{M2h0ijA}
\end{eqnarray}
Here
\begin{eqnarray}
p_S=\frac{i}{k^2-m^2_S+i m_S\,\Gamma_S} \, . \label{pS}
\end{eqnarray}
with the transfer momentum square $k^2$; $p$, $q$, $k_{l_i}$ are
four momentum of incoming $H^0$, outgoing $h^0$, incoming $l_i^-$
lepton respectively. The angle $\theta_{ij}$ carries the
information about the complexity of the Yukawa coupling
$\bar{\xi}^{E}_{N,j i}$ with the parametrization
\begin{equation}
\bar{\xi}^{E}_{N,j i}=|\bar{\xi}^{E}_{N,j i}|\, e^{i \theta_{ij}}
\, , \label{xicomplex}
\end{equation}
Similarly, the matrix element square of the process $H^0
\rightarrow A^0 l_i^- l_j^+$ is obtained as
\begin{eqnarray}
|M|^2=\frac{g^4}{4 m^4_{W}}(A'_1+A'_2+A'_3) \nonumber \\
\label{M2A0ij}
\end{eqnarray}
where
\begin{eqnarray}
A'_1&=&  \frac{1}{2 (m^2_{H^0}+2 p.k_{l_i})^2}\, \, \Bigg\{
m^2_{l_i}\, |\bar{\xi}^{E}_{N,j i}|^2 \, \Bigg( 2
(p.k_{l_i})^2+(m^2_{H^0}-4
m^2_{l_i})\,q.k_{l_i}+p.k_{l_i}\,(m^2_{H^0}\nonumber \\ &+&
4\,m_{l_i}(-m_{l_j}+2\,m_{l_i}+2 m_{l_j}\, sin^2\theta_{l_i l_j})-
2 p.q)+ m_{l_i}(4\, m^2_{l_i}(m_{l_i}-m_{l_j}+2\,m_{l_j}\,
sin^2\theta_{ij})\nonumber\\ &+&
m^2_{H^0}(3\,m_{l_i}-m_{l_j}+2\,m_{l_j}\, sin^2\theta_{l_i
l_j})-4\,m_{l_j}\, p.q \Bigg)\Bigg\} \, ,
\nonumber \\
A'_2&=&\frac{1}{(m^2_{H^0}+2 p.k_{l_i})}\, \, \Bigg\{ 2\,
m_{l_i}\,m^2_{A^0} |\bar{\xi}^{E}_{N,j i}|^2 \,Im[p_{A^0}] \Bigg(
(3
m_{l_i}-m_{l_j}+2\,m_{l_j}\,sin^2\theta_{ij}) p.k_{l_i}\nonumber \\
&+& m_{l_i}\,(m^2_{H^0}+ 2\, m^2_{l_i}-2\, m_{l_i} m_{l_j}+4\,
m_{l_i} m_{l_j}\,sin^2\theta_{ij} -2 \,q.(k_{l_i}-p))\Bigg)
\Bigg\}\, , \nonumber
\\
A_3&=& 2\, m^4_{A^0}\, |\bar{\xi}^{E}_{N,j i}|^2 \,Abs[p_{A^0}]^2
\Bigg(m_{l_i}(m_{l_i}-m_{l_j}+2\,m_{l_j}\,sin^2\theta_{ij})+(p-q).k_{l_i}
\Bigg)\, , \label{M2A0ijA}
\end{eqnarray}
where $q$ is four momentum of outgoing $A^0$. Notice that we use
the parametrization
\begin{equation}
\xi^{E}_{N,ij}= \sqrt{\frac{4 G_F}{\sqrt {2}}}
\bar{\xi}^{E}_{N,ij} \, , \label{ksipar}
\end{equation}
where $G_F=1.6637 \times 10^{-5} (GeV^{-2})$ is the fermi
constant.
Finally, the decay width $\Gamma$ is obtained in the $H^0$ boson
rest frame using the well known expression
\begin{equation}
d\Gamma=\frac{(2\, \pi)^4}{m_{H^0}} \, |M|^2\,\delta^4
(p-\sum_{i=1}^3 p_i)\,\prod_{i=1}^3\,\frac{d^3p_i}{(2 \pi)^3 2
E_i} \,
 ,
\label{DecWidth}
\end{equation}
where $p$ ($p_i$, i=1,2,3) is four momentum vector of $H^0$ boson,
($h^0$ ($A^0$) boson, incoming $l_i^-$, outgoing $l_j^-$ leptons).
\section{Discussion}
In this section, we study the LFV $H^0\rightarrow h^0\, (\tau^-
\mu^+ +\tau^+ \mu^-)$, $H^0\rightarrow A^0\, (\tau^- \mu^+ +\tau^+
\mu^-)$ and the LFC $H^0\rightarrow h^0\, \tau^- \tau^+$,
$H^0\rightarrow A^0\, \tau^- \tau^+$ decays in the model III. The
source of the LFV interactions are the Yukawa couplings
$\xi^{E}_{N,ij}$, which are free parameters of the model used, and
they should be restricted by using the appropriate experimental
measurements. There are various theoretical works to predict the
upper limits of these couplings in the literature. The upper limit
of the coupling $\xi^{E}_{N,\tau\mu}$ has been predicted as $\sim
0.15$, by using the experimental result of anomalous magnetic
moment of muon in \cite{Erilano}. The low energy limits of the LFV
couplings from muon g-2 measurement has been presented in
\cite{Assamagan}. In this work the upper bound of the LFV coupling
$\lambda_{\mu\tau}$ was predicted in the range $1-10$ for type b
2HDM and $10-100$ for type a 2HDM, for various values of Higgs
masses, tan$\beta$ and mixing angle $\alpha$. Here the well known
parametrization $\xi^E_{N,\mu\tau}=\lambda_{\mu\tau}
\frac{\sqrt{m_\tau m_\mu}}{v}$ has been used where the numerical
value of v is $246\, GeV$. The decay width of the LFV
$H^0\rightarrow \tau^- \mu^+$ process in the type b 2HDM has been
predicted as $10^{-2} \lambda^2$ ($10^{-4} \lambda^2$) for
tan$\beta \sim 50$ (tan$\beta \sim 5$) and the small mixing
between CP even neutral Higgs bosons. Furthermore, the numerical
value of the decay width for the LFC $H^0\rightarrow \tau^-
\tau^+$ was estimated as $10^{-1} \lambda^2$ ($10^{-3}
\lambda^2$). \cite{ErHay2} is devoted to the analysis of the LFV
$H^+\rightarrow W^+\, (\tau^- \mu^+ +\tau^+ \mu^-)$ and the LFC
$H^+\rightarrow W^+\, \tau^- \tau^+$ decays in the framework of
the model III. In this work the coupling $\xi^{E}_{N,\tau \mu}$
($\xi^{E}_{N,\tau \tau}$ ) has been estimated as $\sim 0.03\, GeV$
($\sim 0.15\, GeV$) for the range of the decay width $\Gamma
(H^+\rightarrow W^+\, (\tau^- \mu^+ +\tau^+ \mu^-))$  ($\Gamma
(H^+\rightarrow W^+\, \tau^- \tau^+)$) $(10^{-11}-10^{-5})\, GeV$
($(10^{-9}-10^{-4})\, GeV$) and the charged Higgs mass $200 \leq
m_{H^+} \leq 400\, GeV$. Here the Yukawa coupling
$\xi^{E}_{N,\tau\tau}$ plays the main role in the existence of the
LFC decay $H^0\rightarrow h^0 (A^0)\, \tau^- \tau^+$ decay and its
prediction would be possible with  the future experimental
measurement of these decays.

Now we start to analyze the LFV $H^0\rightarrow h^0 (A^0) \,
(\tau^- \mu^+ + \tau^+ \mu^-)$ and LFC $H^0\rightarrow h^0 (A^0)
\, \tau^- \tau^+$ decays. In the numerical calculations we take
the total decay widths of $h^0$ and $A^0$
$\Gamma_{h^0}=\Gamma_{A^0}\sim 0.1\, GeV$, which is at the same
order of magnitude of $\Gamma_{H^0}$.

\begin{figure}[htb]
\vskip -3.0truein \centering \epsfxsize=6.8in
\leavevmode\epsffile{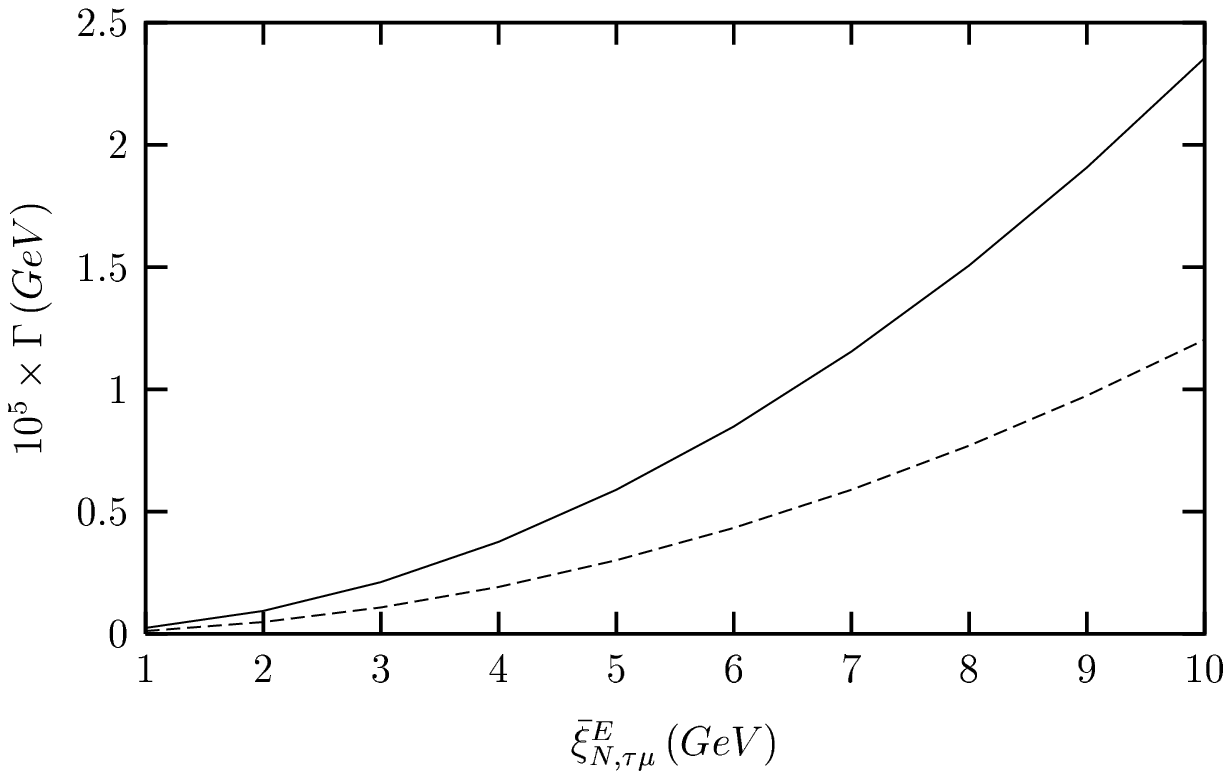} \vskip -3.0truein \caption[]{
$\bar{\xi}^{E}_{N,\tau \mu}$ dependence of the decay width
$\Gamma$ for the decay $H^0\rightarrow h^0 (A^0) \, (\tau^- \mu^+
+ \tau^+ \mu^-)$, for the real coupling $\bar{\xi}^{E}_{N,\tau
\mu}$, $m_{H^0}=150\, GeV$, $m_{h^0}=85\, GeV$ ($m_{A^0}=90\,
GeV$). Here solid (dashed) line represents the case for the output
$h^0$ ($A^0$).} \label{dWLFVtaumu}
\end{figure}

In Fig.\ref{dWLFVtaumu}, we present $\bar{\xi}^{E}_{N,\tau \mu}$
dependence of the decay width $\Gamma$ for the decay
$H^0\rightarrow h^0 (A^0) \, (\tau^- \mu^+ + \tau^+ \mu^-)$, for
the real coupling $\bar{\xi}^{E}_{N,\tau \mu}$, $m_{H^0}=150\,
GeV$, $m_{h^0}=85\, GeV$ ($m_{A^0}=90\, GeV$). Here the solid
(dashed) line represents the case for the output $h^0$ ($A^0$).
The $\Gamma$ is at the order of magnitude of $10^{-6}-10^{-5}$ for
the range of $\bar{\xi}^{E}_{N,\tau \mu}$, $1 \,GeV<
\bar{\xi}^{E}_{N,\tau \mu} < 10 \, GeV $ and it enhances with the
increasing values of the coupling $\bar{\xi}^{E}_{N,\tau \mu}$.
The $\Gamma$ for the output $h^0$ is greater than the one for the
output $A^0$.

\begin{figure}[htb]
\vskip -3.0truein \centering \epsfxsize=6.8in
\leavevmode\epsffile{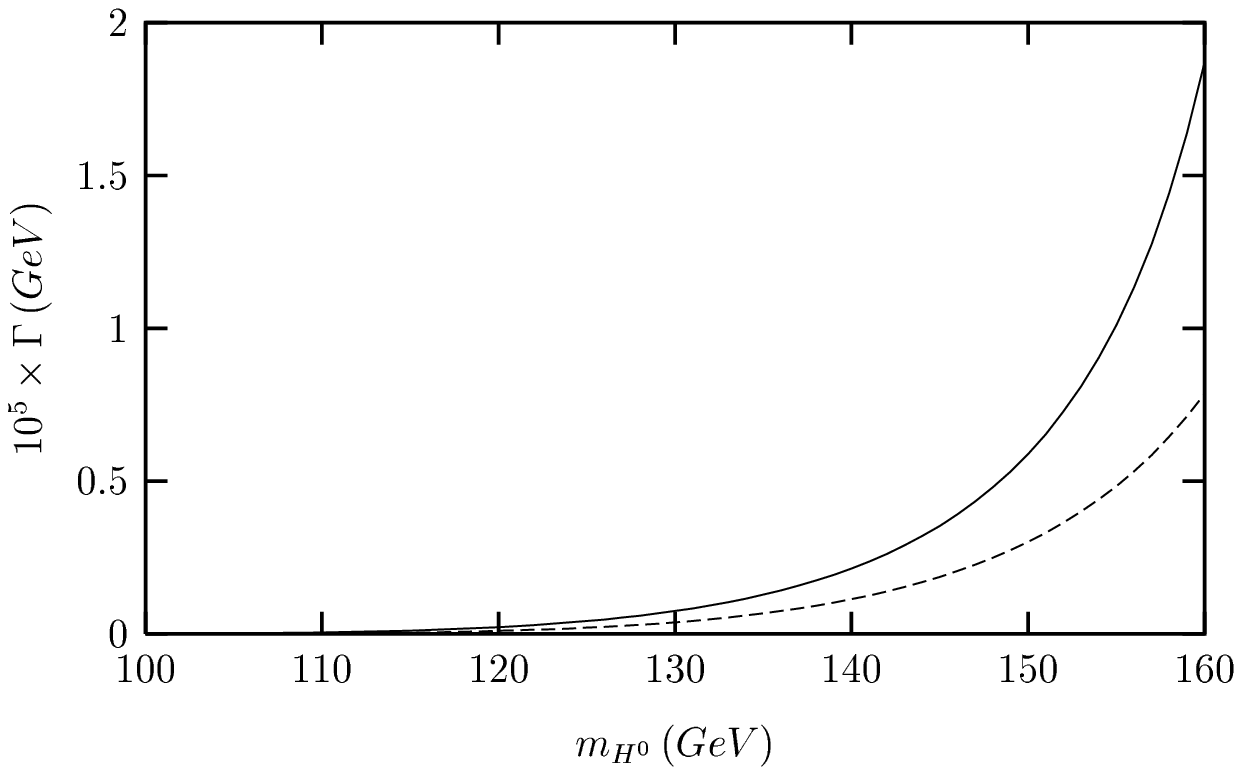} \vskip -3.0truein \caption[]{The
$m_{H^0}$ dependence of the $\Gamma$ for the decay $H^0\rightarrow
h^0 (A^0) \, (\tau^- \mu^+ + \tau^+ \mu^-)$, for the fixed values
of $\bar{\xi}^{E}_{N,\tau \mu}=5\, GeV$, $m_{h^0}=85\, GeV$
($m_{A^0}=90\, GeV$). Here solid (dashed) line represents the case
for the output $h^0$ ($A^0$)} \label{dWLFVmH0}
\end{figure}

Fig. \ref{dWLFVmH0} represents the $m_{H^0}$ dependence of the
$\Gamma$ for the decay $H^0\rightarrow h^0 (A^0) \, (\tau^- \mu^+
+ \tau^+ \mu^-)$, for the fixed values of $\bar{\xi}^{E}_{N,\tau
\mu}=5\, GeV$, $m_{h^0}=85\, GeV$ ($m_{A^0}=90\, GeV$). The
$\Gamma$ is strongly sensitive to the Higgs mass $m_{H^0}$ and it
enhances with the increasing values of the Higgs mass, as
expected.

\begin{figure}[htb]
\vskip -3.0truein \centering \epsfxsize=6.8in
\leavevmode\epsffile{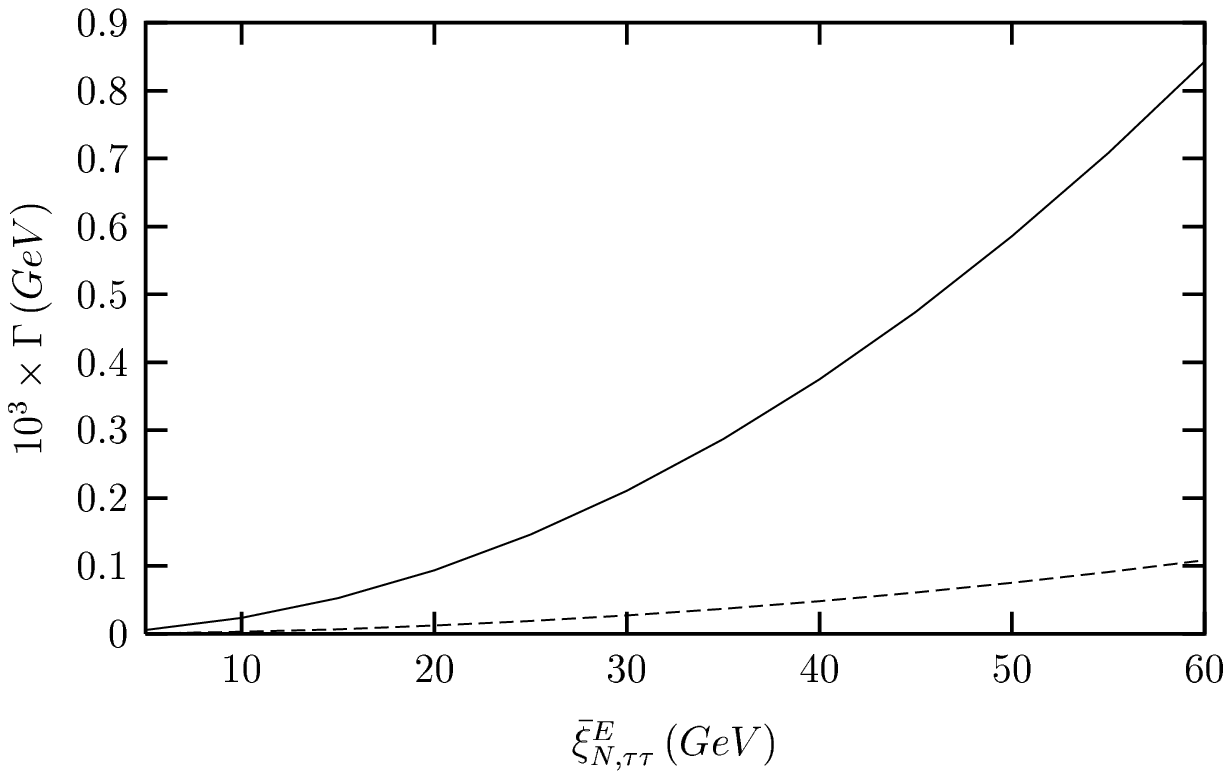} \vskip -3.0truein \caption[]{
$\bar{\xi}^{E}_{N,\tau \tau}$ dependence of the decay width
$\Gamma$ for the decay $H^0\rightarrow h^0 (A^0) \, \tau^-
\tau^+$, for the real coupling $\bar{\xi}^{E}_{N,\tau \tau}$,
$m_{H^0}=150\, GeV$, $m_{h^0}=85\, GeV$ ($m_{A^0}=90\, GeV$). Here
solid (dashed) line represents the case for the output $h^0$
($A^0$).}
 \label{dWLFCtautau}
\end{figure}

Fig.\ref{dWLFCtautau} is devoted to $\bar{\xi}^{E}_{N,\tau \tau}$
dependence of the decay width $\Gamma$ for the decay
$H^0\rightarrow h^0 (A^0) \, \tau^- \tau^+$, for the real coupling
$\bar{\xi}^{E}_{N,\tau \tau}$, $m_{H^0}=150\, GeV$, $m_{h^0}=85\,
GeV$ ($m_{A^0}=90\, GeV$). Here the solid (dashed) line represents
the case for the output $h^0$ ($A^0$). The $\Gamma$ is at the
order of magnitude of $10^{-4}-10^{-3}$ for the range of
$\bar{\xi}^{E}_{N,\tau \tau}$,  $20\, GeV< \bar{\xi}^{E}_{N,\tau
\tau} < 60\, GeV $ and it enhances with the increasing values of
the coupling $\bar{\xi}^{E}_{N,\tau \tau}$. Similar to the LFV
process $H^0\rightarrow h^0 (A^0) \, (\tau^- \mu^+ + \tau^+
\mu^-)$, the $\Gamma$ for the output $h^0$ is greater than the one
for the output $A^0$.

\begin{figure}[htb]
\vskip -3.0truein \centering \epsfxsize=6.8in
\leavevmode\epsffile{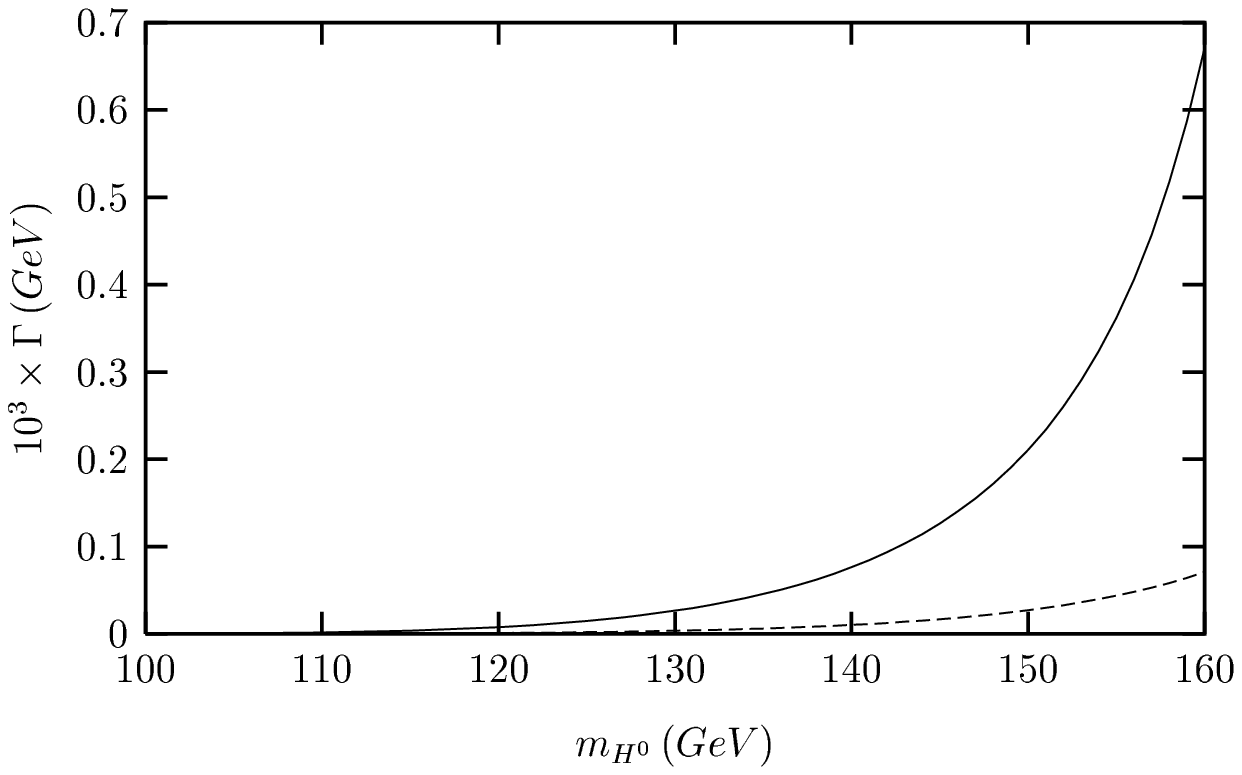} \vskip -3.0truein \caption[]{
$m_{H^0}$ dependence of the $\Gamma$ for the decay $H^0\rightarrow
h^0 (A^0) \, \tau^- \tau^+$, for the fixed values of
$\bar{\xi}^{E}_{N,\tau \tau}=30\, GeV$, $m_{h^0}=85\, GeV$
($m_{A^0}=90\, GeV$). Here solid (dashed) line represents the case
for the output $h^0$ ($A^0$).} \label{dWLFCmH0}
\end{figure}

Fig. \ref{dWLFCmH0} represents the $m_{H^0}$ dependence of the
$\Gamma$ for the decay, for the fixed values of
$\bar{\xi}^{E}_{N,\tau \tau}=30\, GeV$, $m_{h^0}=85\, GeV$
($m_{A^0}=90\, GeV$). The $\Gamma$ is strongly sensitive to the
Higgs mass $m_{H^0}$ and it enhances with the increasing values of
the Higgs mass. It reaches to $\sim 10^{-3} GeV$ for the values of
$m_{H^0}$, $m_{H^0}\sim 160 \, GeV$. The experimental measurement
of this process can give strong clues about the upper limit of the
coupling $\bar{\xi}^{E}_{N,\tau \tau}$ and the Higgs mass
$m_{H^0}$.

\begin{figure}[htb]
\vskip -3.0truein \centering \epsfxsize=6.8in
\leavevmode\epsffile{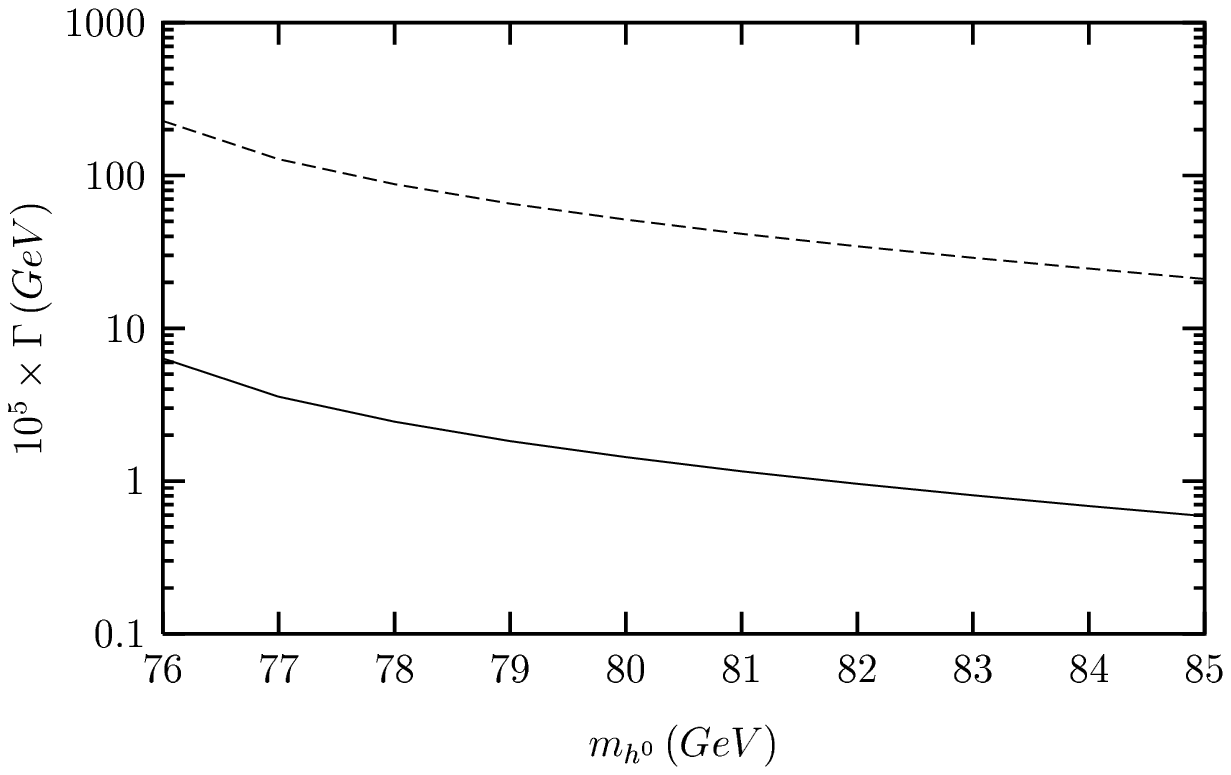} \vskip -3.0truein \caption[]{The
$m_{h^0}$ dependence of the $\Gamma$ of the LFV $H^0\rightarrow
h^0  \, (\tau^- \mu^+ + \tau^+ \mu^-)$ and LFC $H^0\rightarrow h^0
\, \tau^- \tau^+$ decays. The solid (dashed) line represents the
$\Gamma$ of the LFV (LFC) decay, for $\bar{\xi}^{E}_{N,\tau
\mu}=5\, GeV$ ($\bar{\xi}^{E}_{N,\tau \tau}=30\, GeV$),
$m_{H^0}=150\, GeV$.} \label{dWmh0}
\end{figure}

\begin{figure}[htb]
\vskip -3.0truein \centering \epsfxsize=6.8in
\leavevmode\epsffile{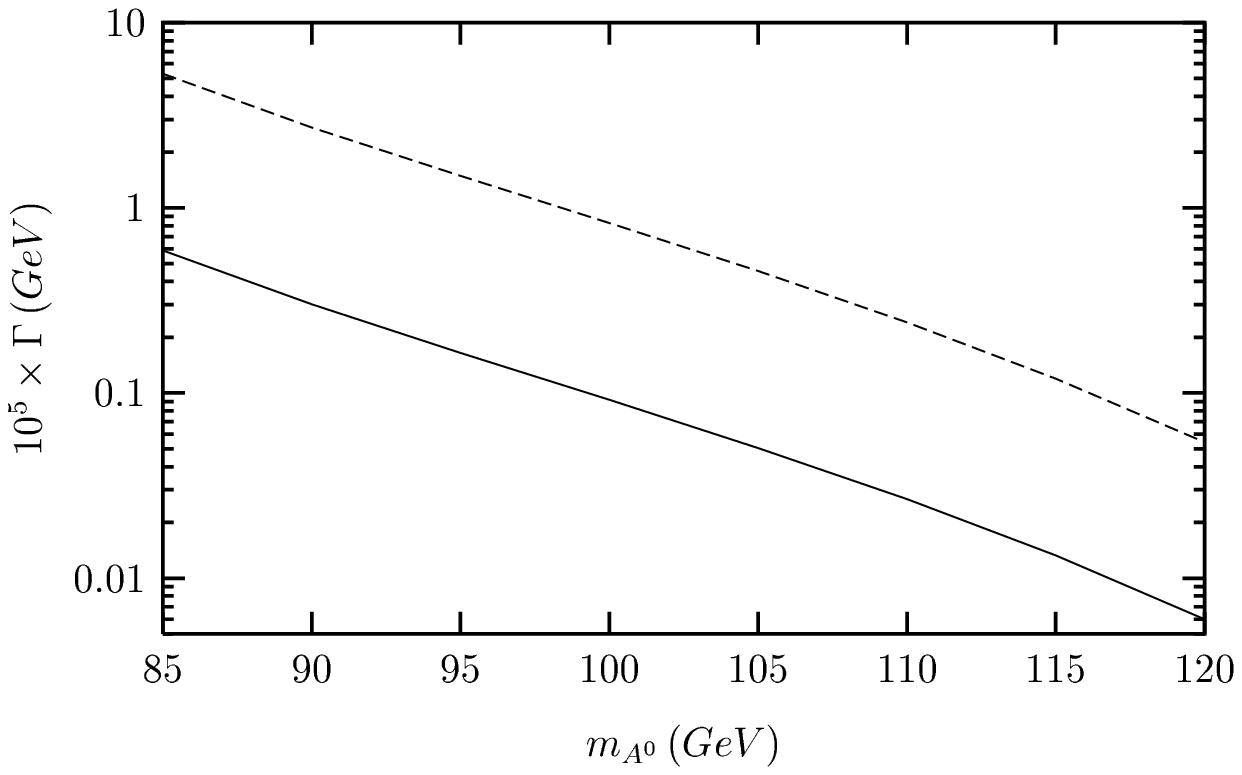} \vskip -3.0truein \caption[]{
$m_{A^0}$ dependence of the $\Gamma$ of the LFV $H^0\rightarrow
A^0 \, (\tau^- \mu^+ + \tau^+ \mu^-)$ and LFC $H^0\rightarrow A^0
\, \tau^- \tau^+$ decays. The solid (dashed) line represents the
$\Gamma$ of the LFV (LFC) decay, for $\bar{\xi}^{E}_{N,\tau
\mu}=5\, GeV$ ($\bar{\xi}^{E}_{N,\tau \tau}=30\, GeV$),
$m_{H^0}=150\, GeV$.}
 \label{dWmA0}
\end{figure}

Finally, we study the $m_{h^0}$ ($m_{A^0}$) dependence of these
LFV and LFC decays and present in the Fig. \ref{dWmh0}
(\ref{dWmA0}). The $m_{h^0}$ dependence of the $\Gamma$ is
presented in Fig. \ref{dWmh0}. The solid (dashed) line represents
the $\Gamma$ of the LFV (LFC) decay, for $\bar{\xi}^{E}_{N,\tau
\mu}=5\, GeV$ ($\bar{\xi}^{E}_{N,\tau \tau}=30\, GeV$),
$m_{H^0}=150\, GeV$. The decay widths increase with the decreasing
values of $m_{h^0}$ and this is informative in the determination
of the Higgs mass $m_{h^0}$. We present the $m_{A^0}$ dependence
of the $\Gamma$ in Fig. \ref{dWmA0}. The solid (dashed) line
represents the $\Gamma$ of the LFV (LFC) decay, for
$\bar{\xi}^{E}_{N,\tau \mu}=5\, GeV$ ($\bar{\xi}^{E}_{N,\tau
\tau}=30\, GeV$), $m_{H^0}=150\, GeV$. Similar to the $m_{h^0}$,
the decay widths increase with the decreasing values of $m_{A^0}$
and it can give a strong clue for the Higgs mass $m_{A^0}$.

Now, we consider the coupling $\bar{\xi}^{E}_{N,ij}$ complex (see
eq. (\ref{xicomplex})) and study the $\sin\,{\theta_{ij}}$
dependence of the decay width. We observe that the decay width is
not sensitive to the complexity of the coupling
$\bar{\xi}^{E}_{N,ij}$.

At this stage we would like to summarize our results:

\begin{itemize}
\item We predict the decay width $\Gamma (H^0\rightarrow h^0
(A^0)\, (\tau^- \mu^+ + \tau^+ \mu^-))$ ($\Gamma (H^0\rightarrow
h^0 (A^0)\, \tau^- \tau^+$)) in the interval $(10^{-6}-10^{-5})\,
GeV$ ($(10^{-4}-10^{-3})\, GeV$), for $100 \,GeV \leq m_{H^0}\leq
160 \,GeV$, at the intermediate values of the coupling
$\bar{\xi}^{E}_{N,\tau \mu}\sim 5 \,GeV$ ($\bar{\xi}^{E}_{N,\tau
\tau}\sim \,30\, GeV$). The future experimental measurement of the
processes under consideration can ensure strong information about
the the upper limit of the coupling $\bar{\xi}^{E}_{N,\tau \mu}$
($\bar{\xi}^{E}_{N,\tau \tau}$) and the Higgs mass $m_{H^0}$.

\item We observe that the decay widths $\Gamma (H^0\rightarrow h^0
(A^0)\, (\tau^- \mu^+ + \tau^+ \mu^-))$ and $\Gamma
(H^0\rightarrow h^0 (A^0)\, \tau^- \tau^+$) are strongly sensitive
to the Higgs masses, $m_{h^0}$ and $m_{A^0}$, respectively. This
observation is useful in the determination of the new Higgs masses
$m_{h^0}$ and $m_{A^0}$.

\item We observe that the decay width $\Gamma (H^0\rightarrow h^0
(A^0)\, (\tau^- \mu^+ + \tau^+ \mu^-))$ ($\Gamma (H^0\rightarrow
h^0 (A^0)\, \tau^- \tau^+$)) is not sensitive to the possible
complexity of the Yukawa coupling.
\end{itemize}

Therefore, the experimental and theoretical analysis of these
decays would ensure strong information about the SM Higgs boson,
and the new physics beyond the SM.

\section{Acknowledgement}
This work has been supported by the Turkish Academy of Sciences in
the framework of the Young Scientist Award Program.
(EOI-TUBA-GEBIP/2001-1-8)

\newpage
\newpage

\begin{thebibliography}{1}
%
\bibitem{Djouadi} A. Djouadi, J. Kalinowski, M. Spira, {\it Comput. Phys.
Commun.} {\bf 108}, 56 (1998).
%
\bibitem{Drollinger} V. Drollinger and T. Muller, D. Denegri,
hep-ph/0111312.
%
\bibitem{Howard} H. E. Haber, M. J. Herrero, H. E. Logan,
S. Penaranda, S. Rigolin, D. Temes {\it Phys. Rev.} {\bf D63},
055004 (2001).
%
\bibitem{RunII} M. Carena, J. S. Conway, H. E. Haber, J. D. Hobbs, et. al.,
Physics at Run II: Supersymmery/Higgs workshop, hep-ph/0010338
(2000).
%
\bibitem{Dittmar} M. Dittmar, H. Dreiner, hep-ph/9703401 (1997);
M. Dittmar, H. Dreiner,{\it Phys. Rev.} {\bf D55}, 167 (1997).

%
\bibitem{Cotti} U. Cotti, L. Diaz-Cruz, C. Pagliarone, E. Vataga,
hep-ph/0111236 (2001).
%
\bibitem{Marfatia} T. Han, D. Marfatia, {\it Phys. Rev. Lett.} {\bf
D86} 1442 (2001).
%
\bibitem{Koerner} J. G. Koerner,A. Pilaftsis, K. Schilcher,
{\it Phys. Rev.} {\bf D47}, 1080 (1993).
%
\bibitem{Assamagan} K. A. Assamagan, A. Deandrea, P.A. Delsart,
hep-ph/0207302 (2002).
%
\bibitem{Erilano} E. Iltan, H. Sundu, hep-ph/0103105 (2001).
%
\bibitem{ErHay2} E. Iltan, H. Sundu, hep-ph/0211343 (2002).
%
\end{thebibliography}
\end{document}